\documentstyle[12pt,epsfig]{article}

\def\bss{\begin{subequations}}
\def\ess{\end{subequations}}

\catcode`\@=11

\newtoks\@stequation

\def\subequations{\refstepcounter{equation}%
  \edef\@savedequation{\the\c@equation}%
  \@stequation=\expandafter{\theequation}
  \edef\@savedtheequation{\the\@stequation}
  \edef\oldtheequation{\theequation}%
  \setcounter{equation}{0}%
  \def\theequation{\oldtheequation\alph{equation}}}

\def\endsubequations{\setcounter{equation}{\@savedequation}%
  \@stequation=\expandafter{\@savedtheequation}%
  \edef\theequation{\the\@stequation}\global\@ignoretrue
  \vspace*{-12pt} \\}

\catcode`\@=11
\def\@citex[#1]#2{%
\if@filesw \immediate \write \@auxout {\string \citation {#2}}\fi
\@tempcntb\m@ne \let\@h@ld\relax \def\@citea{}%
\@cite{%
  \@for \@citeb:=#2\do {%
    \@ifundefined {b@\@citeb}%
      {\@h@ld\@citea\@tempcntb\m@ne{\bf ?}%
      \@warning {Citation `\@citeb ' on page \thepage \space undefined}}%
      {\@tempcnta\@tempcntb \advance\@tempcnta\@ne%
      \@tempcntb\number\csname b@\@citeb \endcsname \relax%
      \ifnum\@tempcnta=\@tempcntb 
        \ifx\@h@ld\relax%
          \edef \@h@ld{\@citea\csname b@\@citeb\endcsname}%
        \else%
          \edef\@h@ld{\ifmmode{-}\else--\fi\csname b@\@citeb\endcsname}%
        \fi%
      \else
        \@h@ld\@citea\csname b@\@citeb \endcsname%
        \let\@h@ld\relax%
      \fi}%
    \def\@citea{,\penalty\@highpenalty\,}%
  }\@h@ld
}{#1}}

\def\@citeb#1#2{{[#1]\if@tempswa , #2\fi}}
\def\@citeu#1#2{{$^{#1}$\if@tempswa , #2\fi }}
\def\@citep#1#2{{#1\if@tempswa , #2\fi}}

\def\bcites{         
        \catcode`\@=11
        \let\@cite=\@citeb
        \catcode`\@=12
}

\def\upcites{         
        \catcode`\@=11
        \let\@cite=\@citeu
        \catcode`\@=12
}

\def\plaincites{      
        \catcode`\@=11
        \let\@cite=\@citep
        \catcode`\@=12
}

\newcount\hour
\newcount\minute
\newtoks\amorpm
\hour=\time\divide\hour by 60
\minute=\time{\multiply\hour by 60 \global\advance\minute by-\hour}
\edef\standardtime{{\ifnum\hour<12 \global\amorpm={am}%
        \else\global\amorpm={pm}\advance\hour by-12 \fi
        \ifnum\hour=0 \hour=12 \fi
        \number\hour:\ifnum\minute<10 0\fi\number\minute\the\amorpm}}
\edef\militarytime{\number\hour:\ifnum\minute<10 0\fi\number\minute}

\def\draftlabel#1{{\@bsphack\if@filesw {\let\thepage\relax
   \xdef\@gtempa{\write\@auxout{\string
      \newlabel{#1}{{\@currentlabel}{\thepage}}}}}\@gtempa
   \if@nobreak \ifvmode\nobreak\fi\fi\fi\@esphack}
        \gdef\@eqnlabel{#1}}
\def\@eqnlabel{}
\def\@vacuum{}
\def\marginnote#1{}
\def\draftmarginnote#1{\marginpar{\raggedright\scriptsize\tt#1}}
\def\draft2label#1{\draftmarginnote{#1} {\@bsphack\if@filesw {\let\thepage\relax
   \xdef\@gtempa{\write\@auxout{\string
      \newlabel{#1}{{\@currentlabel}{\thepage}}}}}\@gtempa
   \if@nobreak \ifvmode\nobreak\fi\fi\fi\@esphack}
        \gdef\@eqnlabel{#1}}
\def\draft{
        \pagestyle{plain}
        \overfullrule=2pt
        \oddsidemargin -.5truein
        \def\@oddhead{\sl \phantom{\today\quad\militarytime} \hfil
        \smash{\Large\sl DRAFT} \hfil \today\quad\militarytime}
        \let\@evenhead\@oddhead
        \let\label=\draftlabel
        \let\marginnote=\draftmarginnote
        \def\ps@empty{\let\@mkboth\@gobbletwo
        \def\@oddfoot{\hfil \smash{\Large\sl DRAFT} \hfil}
        \let\@evenfoot\@oddhead}
        \def\@eqnnum{(\theequation)\rlap{\kern\marginparsep\tt\@eqnlabel}%
        \global\let\@eqnlabel\@vacuum}  }

\def\draft2{
        \pagestyle{plain}
        \overfullrule=2pt
        \oddsidemargin -.5truein
        \def\@oddhead{\sl \phantom{\today\quad\militarytime} \hfil
        \smash{\Large\sl DRAFT} \hfil \today\quad\militarytime}
        \let\@evenhead\@oddhead
        \let\label=\draft2label
        \let\marginnote=\draftmarginnote
        \def\ps@empty{\let\@mkboth\@gobbletwo
        \def\@oddfoot{\hfil \smash{\Large\sl DRAFT} \hfil}
        \let\@evenfoot\@oddhead} }

\def\blackfonts{
        \font\blackboard=msbm10 scaled\magstep1
        \font\blackboards=msbm8
        \font\blackboardss=msbm6
}

\def\nblack{            
        \def\ZZ{{Z \n{10} Z}}
        \def\NN{{N \n{14} N}}
        \def\CC{{C \n{11} C}}
        \def\RR{{R \n{11} R}}
        \def\QQ{{Q \n{12} Q}}
        \def\PP{{P \n{11} P}}
}

\def\prep{         
        \catcode`\@=11
        \input art10.sty
        \catcode`\@=12
        
        \let\small\null
        \def\blackfonts{
                \font\blackboard=msbm10
                \font\blackboards=msbm7
                \font\blackboardss=msbm5
        }
        \let\sl\it
        \twocolumn
        \sloppy
        \voffset=-2.54truecm
        \hoffset=-2.54truecm
        \flushbottom
        \parindent 1em
        \leftmargini 2em
        \leftmarginv .5em
        \leftmarginvi .5em
        \marginparwidth 48pt
        \marginparsep 10pt
        \setlength{\columnsep}{2truecm}
        \setlength{\textwidth}{25.4truecm}
        \setlength{\textheight}{17truecm}
        \baselineskip=16pt
        \oddsidemargin .18truein
        \evensidemargin .17truein
}

\def\eqalign#1{\null\,\vcenter{\openup\jot\m@th
  \ialign{\strut\hfil$\displaystyle{##}$&$\displaystyle{{}##}$\hfil
      \crcr#1\crcr}}\,}
\def\eqalignno#1{\displ@y \tabskip\centering
  \halign to\displaywidth{\hfil$\@lign\displaystyle{##}$\tabskip\z@skip
    &$\@lign\displaystyle{{}##}$\hfil\tabskip\centering
    &\llap{$\@lign##$}\tabskip\z@skip\crcr
    #1\crcr}}

\def\section{\@startsection {section}{1}{\z@}{3.ex plus 1ex minus
 .2ex}{2.ex plus .2ex}{\large\bf}}
\def\subsection{\@startsection{subsection}{2}{\z@}{2.75ex plus 1ex minus
 .2ex}{1.5ex plus .2ex}{\bf}}        

\def\appendix{{\newpage\section*{Appendices}}\let\appendix\section%
        {\setcounter{section}{0}
        \gdef\thesection{\Alph{section}}}\section}

\def\abstract{\if@twocolumn
\section*{Abstract}
\else 
\begin{center}
{\bf Abstract\vspace{-.5em}\vspace{0pt}}
\end{center}
\quotation
\fi}

\catcode`\@=12

\newcommand{\beq}{\begin{equation}}
\newcommand{\eeq}{\end{equation}}
\newcommand{\beqa}{\begin{eqnarray}}
\newcommand{\eeqa}{\end{eqnarray}}

\newcommand{\e}{{\rm e}}

\def\noj#1,#2,{{\bf #1} (19#2)\ }
\def\jou#1,#2,#3,{{\sl #1\/ }{\bf #2} (19#3)\ }
\def\ann#1,#2,{{\sl Ann.\ Physics\/ }{\bf #1} (19#2)\ }
\def\cmp#1,#2,{{\sl Comm.\ Math.\ Phys.\/ }{\bf #1} (19#2)\ }
\def\ma#1,#2,{{\sl Math.\ Ann.\/ }{\bf #1} (19#2)\ }
\def\ng#1,#2,{{\sl Nagoya.\ Math.\ J.\/ }{\bf #1} (19#2)\ }
\def\jd#1,#2,{{\sl J.\ Diff.\ Geom.\/ }{\bf #1} (19#2)\ }
\def\invm#1,#2,{{\sl Invent.\ Math.\/ }{\bf #1} (19#2)\ }
\def\cq#1,#2,{{\sl Class.\ Quantum Grav.\/ }{\bf #1} (19#2)\ }
\def\cqg#1,#2,{{\sl Class.\ Quantum Grav.\/ }{\bf #1} (19#2)\ }
\def\ijmp#1,#2,{{\sl Int.\ J.\ Mod.\ Phys.\/ }{\bf A#1} (19#2)\ }
\def\jmphy#1,#2,{{\sl J.\ Geom.\ Phys.\/ }{\bf #1} (19#2)\ }
\def\jams#1,#2,{{\sl J.\ Amer.\ Math.\ Soc.\/ }{\bf #1} (19#2)\ }
\def\grg#1,#2,{{\sl Gen.\ Rel.\ Grav.\/ }{\bf #1} (19#2)\ }
\def\mpl#1,#2,{{\sl Mod.\ Phys.\ Lett.\/ }{\bf A#1} (19#2)\ }
\def\nc#1,#2,{{\sl Nuovo Cim.\/ }{\bf #1} (19#2)\ }
\def\np#1,#2,{{\sl Nucl.\ Phys.\/ }{\bf B#1} (19#2)\ }
\def\pl#1,#2,{{\sl Phys.\ Lett.\/ }{\bf #1B} (19#2)\ }
\def\pla#1,#2,{{\sl Phys.\ Lett.\/ }{\bf #1A} (19#2)\ }
\def\pr#1,#2,{{\sl Phys.\ Rev.\/ }{\bf #1} (19#2)\ }
\def\prd#1,#2,{{\sl Phys.\ Rev.\/ }{\bf D#1} (19#2)\ }
\def\prl#1,#2,{{\sl Phys.\ Rev.\ Lett.\/ }{\bf #1} (19#2)\ }
\def\prp#1,#2,{{\sl Phys.\ Rept.\/ }{\bf #1C} (19#2)\ }
\def\ptp#1,#2,{{\sl Prog.\ Theor.\ Phys.\/ }{\bf #1} (19#2)\ }
\def\ptpsup#1,#2,{{\sl Prog.\ Theor.\ Phys.\/ Suppl.\/ }{\bf #1} (19#2)\ }
\def\rmp#1,#2,{{\sl Rev.\ Mod.\ Phys.\/ }{\bf #1} (19#2)\ }
\def\yadfiz#1,#2,#3[#4,#5]{{\sl Yad.\ Fiz.\/ }{\bf #1} (19#2) #3%
\ [{\sl Sov.\ J.\ Nucl.\ Phys.\/ }{\bf #4} (19#2) #5]}
\def\zh#1,#2,#3[#4,#5]{{\sl Zh.\ Exp.\ Theor.\ Fiz.\/ }{\bf #1} (19#2) #3%
\ [{\sl Sov.\ Phys.\ JETP\/ }{\bf #4} (19#2) #5]}

\def\beq{\begin{equation}}
\def\eeq{\end{equation}}
\def\beqar{\begin{eqnarray}}
\def\eeqar{\end{eqnarray}}

\newcommand{\be}{\begin{equation}}
\newcommand{\ee}{\end{equation}}
\newcommand{\bea}{\begin{eqnarray}}
\newcommand{\eea}{\end{eqnarray}}

\def\nfrac#1#2{{\displaystyle{\vphantom1\smash{\lower.5ex\hbox{\small$#1$}}%
        \over\vphantom1\smash{\raise.25ex\hbox{\small$#2$}}}}}

\def\d#1{{}_{#1}}

\def\n#1{\mskip-#1mu}

\def\lae{\mathrel{\mathop{\smash{\lower .5 ex \hbox{$\stackrel<\sim$}}}}}
\def\lae{\mathrel{\mathop{\smash{\lower .5 ex \hbox{$\stackrel>\sim$}}}}}

\def\pb{\bar\pa}

\def\l:{\mathopen{:}\,}
\def\r:{\,\mathclose{:}}

\def\ra{\rangle}

\catcode`\@=11
\def\theequation{\arabic{equation}}
\catcode`\@=12

\nblack
\bcites

\nblack

\catcode`\@=11
\def\theequation{\thesection.\arabic{equation}}
\@addtoreset{equation}{section}
\@addtoreset{footnote}{section}
\@addtoreset{footnote}{subsection}
\catcode`\@=12

\def\d{{\rm d}}

\newcommand{\beqn}{\begin{equation}}
\newcommand{\eeqn}{\end{equation}}
\newcommand{\beqnarray}{\begin{eqnarray}}
\newcommand{\eeqnarray}{\end{eqnarray}}

\newcommand{\bv}{\bar{v}}
\newcommand {\bear} [1] {\begin {array} {#1}}
\newcommand {\ear} {\end {array}}

\newcommand {\beqarn} {\begin{eqnarray*}}
\newcommand {\eeqarn} {\end{eqnarray*}}

\def\nonu{\nonumber}
\def\diff{\partial}
\def\diffb{\bar{\partial}}
\def\diffp{\partial_+}
\def\diffm{\partial_-}
\def\a{\alpha}
\def\t{\tau}
\def\s{\sigma}
\def\cl{{\cal L}}
\def\ch{{\cal H}}
\def\gh{\hat{\Gamma}}
\def\g{\Gamma}

\def\delh{\hat{\nabla}}

\def\delhp{\hat{\nabla}^+}
\def\delhm{\hat{\nabla}^-}
\def\lb{\bar{L}}
\def\lt{\tilde{L}}
\def\lbt{\tilde{\bar{L}}}
\def\pb{\bar{P}}

\def\tp{T_{++}}
\def\tm{T_{--}}
\def\tt{\tilde{T}}
\def\ttp{\tilde{T}_{++}}
\def\ttm{\tilde{T}_{--}}
\def\d{\delta}
\def\ppf{\frac{1}{4 \pi \a'}}
\def\ppe{\frac{1}{8\pi \a'}}
\def\bb{\bar{B}}
\def\difflr{\stackrel{\leftrightarrow}{\partial}}
\def\e{\epsilon}
\def\ba{\bar{\alpha}}
\def\bbe{\bar{\epsilon}}
\def\bx{\bar{\xi}}
\def\bv{\bar{v}}
\def\bd{\bar{\delta}}
\def\tr{\rm Tr}
\def\ra{\rightarrow}
\def\L{\Lambda}

\def\omhp{\hat{\omega}^+}
\def\omhm{\hat{\omega}^-}

\begin{document}

\begin{titlepage}

\begin{center}
\today
\hfill LBNL-41327, UCB-PTH-98/08\\
\hfill           hep-th/9802022

\vskip 1.5 cm
{\large \bf New Spin-Two Gauged Sigma Models and\\
 General Conformal Field Theory}
\vskip 1 cm 
{J. de Boer and M. B. Halpern}\\
\vskip 0.5cm
{\sl Department of Physics,
University of California at Berkeley\\
366 Le\thinspace Conte Hall, Berkeley, CA 94720-7300, U.S.A.\\
and\\
Theoretical Physics Group, Mail Stop 50A--5101\\
Ernest Orlando Lawrence Berkeley National Laboratory\\
Berkeley, CA 94720, U.S.A.\\}

\end{center}

\vskip 0.5 cm
\begin{abstract}

Recently, we have studied the general Virasoro construction  at one loop
in the background of the general non-linear sigma model. Here, we 
find the action formulation of these new conformal field theories 
when the background
sigma model is itself conformal. In this case, the new conformal
field theories are described by a large class of new spin-two gauged 
sigma models. As examples of the new actions, we discuss the spin-two
gauged WZW actions, which describe the conformal field theories of
the generic affine-Virasoro construction, and the spin-two gauged 
$g/h$ coset constructions. We are able to identify the latter
as the actions of the local Lie $h$-invariant conformal 
field theories, a large
class of generically irrational conformal field theories 
with a local gauge symmetry.

\end{abstract}

\end{titlepage}

\section{Introduction}

The general affine-Virasoro construction \cite{r1,r2,r3} 
describes the general conformal
stress tensor 
\be
\label{y0}
T=L^{ab} :J_a J_b: , \qquad a,b=1,\ldots,\dim\,g
\ee
in the background of the WZW action on Lie $g$, where $J_a$
are the currents of affine Lie $g$ \cite{r4,r5,r6}, and the coefficients
$L^{ab}$ satisfy the Virasoro master equation. The conformal
field theories (CFTs) described by (\ref{y0}) have generically irrational
central charge, even when the theories are unitary. The generic
affine-Virasoro action \cite{r7,r8,r9} is a large set of spin-two gauged
WZW actions which describe the generic CFT whose stress
tensors have the form (\ref{y0}). The spin-two nature of the
generic theory is a consequence of $K$-conjugation covariance 
\cite{r6,r10,r11,r1},
which tells us that each $T$ comes with a commuting $K$-conjugate 
partner $\tt$
\be
\label{y1}
K_g: \qquad T_g = T+\tt
\ee
such that $T$ and $\tt$ sum to the affine-Sugawara construction
\cite{r6,r10,r12,r13} $T_g$ on $g$. See \cite{r3} for a review of the
general affine-Virasoro construction and irrational conformal field
theory.

Recently, the first part of this program has been extended \cite{r14,r15} 
from the WZW background to the
general non-linear sigma model. In particular, the general conformal
stress tensor 
\be
\label{y2}
T\sim L_{ij} \diffp x^i \diffp x^j , \qquad i,j=1,\ldots,\dim\, M
\ee
was studied at one loop in the background of the general non-linear
sigma model, including the dilaton, and a unified Einstein-Virasoro
master equation was obtained for the coefficient $L_{ij}$. It 
remains therefore to find the actions of the new CFTs associated 
to these more general stress tensors.

In this paper, we confine ourselves to those background sigma models
which are themselves conformal, in which case the more general 
stress tensors (\ref{y2}) also
exhibit $K$-conjugation covariance in the form
\be 
\label{y3}
K_G: \qquad T_G=T+\tt
\ee
where $T_G$ is the conformal stress tensor of the background sigma model.

Following \cite{r7}, 
we find that the generic conformal field theory of this type
is described by a very large class of 
generically new
spin-two gauged sigma models, including
the general non-linear sigma model itself,
the Polyakov-gauged general non-linear sigma model,
 and the spin-two gauged WZW actions
as special cases.

As another special case, we study the spin-two gauged coset constructions, 
for which
$K$-conjugation takes the form
\be
\label{y4} 
K_{g/h}: \qquad T_{g/h} = T+\tt
\ee
where $T_{g/h}$ is the stress tensor of the $g/h$ coset 
construction \cite{r6,r10,r11}.
We are able to identify the spin-two gauged coset constructions
as the actions of the local Lie $h$-invariant CFTs \cite{r16}, which
are known to exhibit $K$-conjugation through the coset constructions.
The set of local Lie $h$-invariant CFTs is the large class
of generically irrational CFTs in the Virasoro master equation
with an extra Lie $h$ gauge symmetry, including the coset
constructions as the simplest case. Because of their extra
Lie $h$ gauge symmetry, the actions of this class of 
theories were previously unknown. With the answer in
hand, however, we find that these theories may equivalently
be described as spin-two and spin-one gauged WZW actions.

 More generally, we expect that the class of CFTs described by
the spin-two gauged sigma models is vast: one hint in this direction
is the observation of many other kinds \cite{r16} of $K$-conjugation
covariance beyond the examples $K_g$ and $K_{g/h}$
discussed explicitly here.

\section{Spin-Two Symmetries of the General Sigma Model}

In this section, we review the classical form of the
general non-linear sigma model and its
spin-two symmetries, following the development and notation of
\cite{r14,r15}. 

On any manifold $M$, the Minkowski-space form of the general
non-linear sigma model is
\bss
\bea
\label{eq1}
& S_G= \int d^2 \xi {\cal L}_G & 
\\{}
\label{eq2}
& {\cal L}_G = \frac{1}{8\pi \alpha'} (G_{ij} + B_{ij} ) \diffp x^i \diffm x^j
& \\{}
\label{eq3}
& d^2 \xi = d\t d\s, \qquad  \diff_{\pm} = \diff_{\t} \pm \diff_{\s}
& \\{}
\label{n1}
& H_{ijk} = \diff_i B_{jk} + \diff_j B_{ki} + \diff_k B_{ij}
& 
\eea \label{u51}
\ess
where $0\leq \s \leq 2\pi$,
$x^i$, $i=1,\ldots,\dim\, M$ are coordinates on $M$ and
$\a'$ is the string tension. The fields $G_{ij}$ and $H_{ijk}$ 
are respectively the metric and the torsion field on $M$. The
equations of motion can be written in two equivalent forms
\bss
\bea
\label{eq4}
& 
\diffp \diffm x^i + \diff_{\pm} x^j \diff_{\mp} x^k \gh^{\mp i}_{jk} =0
& \\{}
\label{eq5}
& \gh^{\pm i}_{jk} = \g_{jk}{}^i \pm \frac{1}{2} H_{jk}{}^i
& 
\eea
\ess
where $\hat{\g}^{\pm}$ are the generalized Christoffel connections with
torsion.

We also introduce the vielbein $e_i{}^a$, $a=1,\ldots,\dim\, M$ on $M$,
\bss
\bea
\label{v1}
& G_{ij} = e_i{}^a G_{ab} e_j{}^b & 
\\{}
\label{v2} 
& \delh_i^{\pm} e_j{}^a = \diff_i e_j{}^a -\hat{\g}^{\pm\, k}_{ij} e_k{}^a
+ e_j{}^b (\hat{\omega}^{\pm}_i)_b{}^a = 0 & 
\\{}
\label{v3}
& (\hat{\omega}^{\pm}_i)_a{}^b = 
(\omega_i)_a{}^b \pm \frac{1}{2} H_{ia}{}^b 
& 
\eea
\ess
where $G_{ab}$ is the tangent-space metric and $\hat{\omega}^{\pm}$
are the generalized spin connections with torsion. This gives
the tangent-space form of the equations of motion
\bss
\bea
\label{te1}
& \diff_{\mp} J^{\pm}_a = (\hat{\omega}^{\pm c})_a{}^b 
J_b^{\pm} J_c^{\mp} & 
\\{}
\label{te2}
& J_a^{\pm} = G_{ab} e_i{}^b \diff_{\pm} x^i
& 
\eea \label{u53}
\ess
in terms of the currents $J^{\pm}$.

To obtain the Hamiltonian formulation, we need the Poisson brackets of the
coordinates with the momenta $p_i=\partial \cl_G/\partial \dot{x}^i$,
\bss
\bea
\label{eq6}
& [x^i(\s),p_j(\s')] = i\delta_j^i \delta(\s-\s') & 
\\{}
%
%
& J_a^{\pm} =  e_a{}^i (4\pi\a' p_i + (B_{ij} \pm G_{ij} ) \partial_{\sigma} x^j)
& \label{eq9}
\eea \label{e25}
\ess
and the equal-time current algebra \cite{r14}
\bss
\bea
[J^+_a(\s),J^+_b(\s')] & = & 8\pi i \a' G_{ab} \diff_{\s} \delta(\s-\s')
 + 4 \pi i \a' \delta(\s-\s') 
[  J^-_c \hat{\omega}^{+c}{}_{ab} -J^+_c \hat{\tau}^{+c}{}_{ab} ]
\nonu \\{}
[J^-_a(\s),J^-_b(\s')] & = & -8\pi i \a' G_{ab} \diff_{\s} \delta(\s-\s')
 + 4 \pi i \a' \delta(\s-\s') [J^+_c \hat{\omega}^{-c}{}_{ab}
 - J^-_c \hat{\tau}^{-c}{}_{ab} ]
\nonu \\{}
[J^+_a(\s),J^-_b(\s')] & = & 
  4 \pi i \a' \delta(\s-\s') [  \hat{\omega}^{+}_{ba}{}^c J^+_c
 -   \hat{\omega}^{-}_{ab}{}^{c} J^-_c ]
\label{eq10} \\{}
(\hat{\tau}^{\pm})_{cab} & \equiv&  \omega_{cab} + \omega_{abc} + \omega_{bca}
\pm \frac{1}{2} H_{cab}
\label{eq11}
\eea
\ess
which follows from (\ref{e25}). The sigma model Hamiltonian is
\bss
\bea
\label{eq12}
& H_G  =  \int_0^{2\pi} d\s \ch_G &
\\{}
\label{eq15}
& \ch_G  =   T_{++}^G + T_{--}^G & 
\\{}
& T_{\pm\pm}^G  =  \frac{1}{8\pi\a'} L_G^{ab}
J_a^{\pm} J_b^{\pm}, \qquad L_G^{ab}  =  \frac{G^{ab}}{2} &
\label{n2}
\\{}
\label{eq14} &
\dot{A}  =  i[H_G,A] &
\eea
\ess
where $G^{ab}$ is the inverse of the tangent space metric.

The stress tensors $T_{++}^G$ and $T_{--}^G$ are respectively chiral
and antichiral
\be
\label{n4}
\diff_{\mp} T_{\pm\pm}^G=0
\ee
and satisfy the commuting Virasoro algebras
\bea
[T_{\pm\pm} (\s) , T_{\pm\pm} (\s') ] & = & 
\pm i[T_{\pm\pm}(\s) + T_{\pm\pm}(\s') ]
\partial_{\s} \delta(\s-\s')  \nonu  \\{}
[T_{++}(\s), T_{--}(\s')  ] & = & 0
\label{eq17}
\eea
at equal time. The semiclassical limit of the central charge of
$T_{\pm\pm}^G$ is $c_G=\dim\,M$. 

We turn now to the contruction of new chiral and antichiral stress tensors
in the background of the sigma model. With \cite{r14,r15}, we suppose that the
manifold $M$ supports two covariantly-constant second-rank symmetric
tensors $L$ and $\bar{L}$
\bss
\bea 
& \delhp_i L^{ab} = \delhm_i \lb^{ab}=0 &
\label{eq18}
\\{}
%
%
%
\label{eq21}
& L^{ab} = 2 L^{ac} G_{cd} L^{db}, \qquad 
\lb^{ab} = 2 \lb^{ac} G_{cd} \lb^{db} & 
\eea \label{e210}
\ess
called the (inverse) inertia tensors on $M$. 
The relations (\ref{eq21}) are familiar as the high-level
or semiclassical 
form \cite{r17,r18,r7} of the Virasoro master equation, and are
easily solved as
\be
L_a{}^b=\frac{P_a{}^b}{2},
\qquad \lb_a{}^b = \frac{\bar{P}_a{}^b}{2}
\ee
where $P$ and $\pb$ are orthogonal
projectors. The relations (\ref{eq18})
are called the covariant-constancy conditions; necessary
and
sufficient conditions for the existence of solutions to these
conditions are
known \cite{r14,r15}, and examples will be discussed below.

For each such pair of inertia tensors on $M$, one has the associated
chiral and antichiral stress tensors
\bss
\bea
& \tp  =  \frac{1}{8\pi\a'} L^{ab} J_a^+ J_b^+, \qquad
\tm  =  \frac{1}{8\pi\a'} \lb^{ab} J_a^- J_b^-  &
\label{eq27}
\\{}
\label{n6}
& \diff_{\mp} T_{\pm\pm} = 0 &
\eea\ess
which also
satisfy the two commuting Virasoro algebras in (\ref{eq17}).

The $K$-conjugate \cite{r6,r10,r11,r1} stress tensors
\bss
\bea
\label{n7}
& \ttp = \frac{1}{8\pi\a'} \lt^{ab} J^+_a J^+_b, \qquad
\ttm  = \frac{1}{8\pi\a'} \lbt^{ab} J^-_a J^-_b & 
\\{}
\label{n8}
& \lt^{ab} = L_G^{ab} - L^{ab}, \qquad 
\lbt^{ab} = L_G^{ab} - \lb^{ab} & 
\\{}
\label{n9}
& \diff_{\mp} \tt_{\pm\pm}=0 & 
\eea\ess
are also respectively chiral and antichiral 
and satisfy the Virasoro algebra (\ref{eq17})
because the
corresponding relations
\bss 
\bea 
& \delhp_i \lt^{ab} = \delhm_i \lbt^{ab}=0 & 
\label{eq19}
\\{}
\label{eq22}
& \lt^{ab} = 2 \lt^{ac} G_{cd} \lt^{db}, \qquad 
\lbt^{ab} = 2 \lbt^{ac} G_{cd} \lbt^{db}, &
\eea\ess
follow from (\ref{eq18}), (\ref{eq21}) and (\ref{n8}). 

Counting the
$K$-conjugate stress tensors, this gives us a total of four
commuting Virasoro generators
\be
\label{n10}
\tp,\ttp,\tm,\ttm
\ee
which sum in $K$-conjugate pairs to the sigma
model stress tensors $T^G_{\pm\pm}$
\be
\label{n11}
K_G: \qquad T^G_{\pm\pm} = T_{\pm\pm} + \tt_{\pm\pm} .
\ee
The interpretation of this structure, well-known from the
general affine-Virasoro construction, is that the background
sigma model $T^G_{\pm\pm}$ factorizes into two commuting
$K$-conjugate conformal field theories
\bea
L\,\,{\rm theory} & :  & \tp,\tm 
\nonu \\{}
\lt \,\,{\rm theory} & : & \ttp,\ttm
\label{n12}
\eea
each of which possesses its own chiral and antichiral stress tensors.

The four Virasoro generators $\tp,\ttp,\tm,\ttm$ correspond to spin-two
symmetries of the sigma model action $S_G$ in (\ref{eq1}) and
these generators
can be obtained from $S_G$ by Noether's theorem,
using the symmetries
%
%
%
%
%
%
%
\bea
\delta x^i & = & 2 \xi(\t^+)  L^i{}_j \diffp x^j 
\nonu \\{}
\tilde{\delta} x^i & = & 2 \tilde{\xi}(\t^+)  \lt^i{}_j \diffp x^j 
\nonu \\{}
\bar{\delta} x^i & = & 2 \bar{\xi}(\t^+)  \lb^i{}_j \diffm x^j 
\nonu \\{}
\tilde{\bar{\delta}}  x^i & = & 2 \tilde{\bar{\xi}}(\t^+)  \lbt^i{}_j \diffm x^j 
\nonu \\{}
\t^{\pm} & =  & \t \pm \s
\nonu \\{}
\delta S_G & = & 
\tilde{\delta} S_G\,\, =\,\,
\bar{\delta} S_G\,\, = \,\,
\tilde{\bar{\delta}} S_G\,\, =\,\, 0
\label{eq29}
\eea
where $\xi,\tilde{\xi},\bx,\tilde{\bx}$ are the infinitesimal 
parameters of the transformations.

In \cite{r14}, the one-loop quantum corrections to this classical discussion
were considered in detail, including the dilaton $\Phi$. It was shown there that
this picture, including the $K$-conjugate pairs of Virasoro operators,
survives at one loop\footnote{The classical picture is of course exact in
the quantum theory when the 
background sigma model is the WZW action, since this case corresponds
to the general affine-Virasoro construction.} as long as the background
sigma model is itself conformal ($G$, $B$ and $\Phi$ satisfy the Einstein
equations). In what follows, the discussion applies only to this case.
We also emphasize that the classical inverse inertia tensors $L,\lb,\lt,\lbt$
are only the semiclassical limits of the one-loop inertia tensors of
\cite{r14}, but this is all that is needed \cite{r7} to construct the classical
Hamiltonian and action formulation of the new CFTs. The semiclassical limits
of the central charges of the various stress tensors are
\bss
\bea
& c(T_{++})={\rm rank}\,L, \qquad
c(T_{--})={\rm rank}\,\lb & \\{}
& c(\tt_{++})={\rm rank}\,\lt , \qquad
c(\tt_{--})={\rm rank}\,\lbt & \\{}
& c_G=\dim\, M=c(T_{++}) + c(\tt_{++})= c(T_{--}) + c(\tt_{--}) . &
\eea
\ess
See \cite{r13} for further details 
at the one-loop level.

\section{Hamiltonian and Action Formulations of the New CFT's}

In \cite{r14} and Section~2, the discussion was limited to the construction
of the stress tensors of the new CFTs, and we now want to find the classical 
Hamiltonian and action
formulations of the new theories, following the development of \cite{r7} for
the generic affine-Virasoro action. In what follows, we focus on the $L$ theory,
with stress tensors $T_{\pm\pm}$, but the corresponding 
constructions for the $\lt$ theory follow by
$K$-conjugation
\be \label{n15}
L \leftrightarrow \lt, \qquad \lb \leftrightarrow \lbt,\qquad T_{\pm\pm} \leftrightarrow 
\tt_{\pm\pm}
\ee
at any stage of the discussion.

\subsection{Hamiltonian Formulation}

We begin with the basic Hamiltonian of the $L$ theory
\bss
\bea
\label{eq30}
& H_0 = \int d\s \ch_0
& \\{}
\label{eq31}
& \ch_0 = \tp + \tm
& \\{}
\label{eq32}
& \tp  =  \frac{1}{8\pi\a'} L^{ab} J_a^+ J_b^+, \qquad
\tm  =  \frac{1}{8\pi\a'} \lb^{ab} J_a^- J_b^-
& \\{}
\label{eq33}
& J_a^{\pm} = e_a{}^i (4\pi\a' p_i + (B_{ij} \pm G_{ij}) \diff_{\s} x^j)
& \\{}
\label{eq34}
& \dot{A} = i[H_0,A]
& \\{}
\label{eq34a}
& \diff_{\mp} T_{\pm\pm}=0
& \eea
\ess
where the metric $G_{ij}$, vielbein $e_i{}^a$ and antisymmetric tensor field $B_{ij}$
are those of the background conformal sigma model $S_G$.
Note that the currents $J^{\pm}$
are now defined by their canonical construction (\ref{eq33}), which guarantees
the general current algebra (\ref{eq10}), (\ref{eq11}), although the relation (\ref{te2})
no longer holds (except for the special case $L^{ab}=\lb^{ab}=
L_G^{ab}$, which returns us to the original sigma model). As a
consequence, all four stress tensors $T_{\pm\pm},\tt_{\pm\pm}$
satisfy commuting Virasoro algebras (as they did in the original sigma model)
and we find in particular that
the $K$-conjugate stress tensors (associated to the $\lt$ theory)
are local symmetries of the basic
Hamiltonian
\be
\label{eq35}
 [H_0,\ttp(\s)]=[H_0,\ttm(\s)]=0 .
\ee
This identifies the system as a spin-two gauge theory, and we may follow
Dirac to construct the full Hamiltonian of the $L$-theory
\bss
\bea
\label{eq36}
H& =& \int d\s\, \ch
\\{}
\label{eq37}
\ch & = & \ch_0 + v \cdot \tt
\\{}
& = & \tp + \tm + v \ttp + \bar{v} \ttm
\label{eq38} \\{}
& = & \ch_G + \frac{1}{8\pi \a'} [(v-1) 
\lt^{ab} J_a^+ J_b^+ + (\bar{v}-1) 
 \lbt^{ab} J_a^- J_b^- ]
\label{eq39}
\\{}
\label{eq40}
\dot{A} & = & i[H,A]
\\{}
\diff_{\mp} T_{\pm\pm} & = & 0
\label{eq41}
\eea\ess
in which the $L$-theory is gauged by its $K$-conjugate partner,
the $\lt$ theory. Here, $v$ and $\bar{v}$ are Lagrange multipliers
which form a world-sheet spin-two gauge field \cite{r7}
\be
\label{n16}
\sqrt{-\tilde{h}} \tilde{h}^{mn} = \frac{2}{v+\bar{v}}
\left( \begin{array}{cc} -1 & \frac{1}{2}(v-\bar{v}) \\
\frac{1}{2}(v-\bar{v}) & v\bar{v} \end{array} \right),\qquad
m=(\tau,\sigma)
\ee
called the $K$-conjugate metric. The Hamiltonian (\ref{eq30}) is
correct for the generic $L$ theory, but must be further gauged
if $H_0$ possesses further local symmetries (see Sections~5.3 and~7).

This system possesses a ${\rm Diff}\,S^1 \times\, {\rm Diff}\,S^1$
symmetry generated by the $K$-conjugate stress tensors
\bss
\bea
\label{eq42}
& \partial A  =  i[\int d \s\, \epsilon(\s) \ttp(\s) + \bar{\epsilon}(\s) \ttm(\s) ,A]
& \\{}
\label{eq43}
& \delta v = \epsilon \stackrel{\leftrightarrow}{\partial}_{\s} v, \quad
\delta \bar{v} = -\bar{\epsilon}
 \stackrel{\leftrightarrow}{\partial}_{\s} \bar{v}
& \\{}
\label{eq45}
& \d x^i = e^i{}_a (\epsilon \lt^{ab} J^+_b + \bar{\epsilon} \lbt^{ab} J^-_b)
& \\{}
\label{eq44}
& \d H = \d H_0 = 0
& 
\eea\ess
which is extended to ${\rm Diff}\,S^2$ on passage to the non-linear
form of the action in the usual manner \cite{r7}. 
We shall return to the non-linear
form of the action below.

Here we follow \cite{r8}, going
directly to the linear form of the action via the introduction of
the auxiliary fields $B,\bb$,
\bss\bea
\ch' & = & \ch_G + \ppf [ \a \lt^{ab} B_a B_b + \frac{1}{2}
 (B_a-J_a^+) G^{ab} (B_b-J_b^+) \nonu
\\{}
& & \qquad \qquad 
\quad + \bar{\alpha} \lbt^{ab} \bar{B}_a \bar{B}_b
+ \frac{1}{2} (\bar{B}_a-J^-_a) G^{ab} (\bar{B}_b - J_b^-) ]
\label{eq46} \\{}
\a & =& \frac{1-v}{1+v}, \qquad \bar{\a} = \frac{1-\bar{v}}{1+\bar{v}}
\label{eq47} \\{}
\dot{A} & = & i[H',A] 
\label{eq47a} \\{}
\label{eq48}
\ch'_{(B,\bar{B})_{\ast}} & = & \ch .
\eea\ess
The new Hamiltonian $H'=\int d\s\, \ch'$ is equivalent to the Hamiltonian
$H$, as shown, after using the $(B,\bb)=(B,\bb)_{\ast}$ equations of motion.

\subsection{Action Formulation: The New Spin 2 Gauged Sigma Models}

Defining $\dot{x}^i$ by (\ref{eq47a}) as usual leads to the useful
identities
\bss
\bea
\label{eq49}
J_a^+ + J_a^-&  = & 2[B_a + \bar{B}_a + G_{ab} e_i{}^b \dot{x}^i ]
\\{}
\label{eq50}
J_a^+ - J_a^-  & = &  2 e_a{}^i G_{ij} \diff_{\s} x^j
\\ {}
\label{eq51}
J_a^{\pm} & = & B_a  + \bar{B}_a - G_{ab} e_i{}^b \diff_{\pm} x^i
\eea
\ess
and then, with $\cl'=\dot{x}^i p_i - \ch'$, we find the linear form of the action
of the $L$-theory
\bss
\bea
\label{eq52}
S'& =& \int d^2 \xi\, \cl'
\\{}
\cl' & = & \cl_G + \ppf [\alpha \lt_{ij} B^i B^j +
\bar{\alpha} \lbt_{ij} \bar{B}^i \bar{B}^j 
\nonu \\{}
& & \qquad \qquad
 \quad -(B^i-\diffp x^i) G_{ij} (\bar{B}^j - \diffm x^j) ]
\label{eq54}
\eea \label{e39}
\ess
as a generically new spin-two gauged sigma model. 
This is the central result of this paper.
In fact, the construction describes a very large class of 
CFTs, one family for each pair $\lt,\lbt$
of covariantly constant inertia tensors
\bss
\bea
& \delhp_i \lt_j{}^k = \delhm_i \lbt_j{}^k =0 & 
\label{q1}
\\{}
& \lt_i{}^j = 2 \lt_i{}^k \lt_k{}^j , \qquad
\lbt_i{}^j = 2 \lbt_i{}^k \lbt_k{}^j & 
\label{q2}
\eea \label{qqqq}
\ess
in the background of the original sigma model $S_G$
(recall that (\ref{e210}) is equivalent to (\ref{qqqq})). Necessary and 
sufficient conditions for the solution of (\ref{qqqq}) are given
in \cite{r14,r15} and these references also discuss the conformal
stress tensors of these theories at the one-loop level including
the dilaton.

This action exhibits a spin-two gauge symmetry, or ${\rm Diff}\, S^2$
invariance
\bss
\bea
\d\a  & = & -\diffm\xi + \xi \difflr_+ \a
\label{eq57}
\\{}
\d x^i & = & \xi^i = 2 \xi \lt^i{}_j B^j
\label{eq58}
\\{}
\d B^i & = & \diffp \xi^i - (B^j - \diffp x^j) \xi^k (\gh^+)_{jk}{}^i
\label{eq59}
\\{}
\d \bar{B}^i & = & -\xi^j \bb^k (\gh^-)_{jk}{}^i
\label{eq60}
\eea
\ess\bss
\bea
\bd\ba  & = & -\diffp\bx + \bx \difflr_- \ba
\label{eq62}
\\{}
\bd x^i & = & \bx^i = 2 \bx \lbt^{i}{}_j B^j
\label{eq63}
\\{}
\bd B^i & =  & -\bx^j B^k (\gh^+)_{jk}{}^i
\label{eq64}
\\{}
\bd \bar{B}^i & = &\diffm \bx^i - (\bb^j - \diffm x^j) \bx^k (\gh^-)_{jk}{}^i
\label{eq65}
\eea\ess
\be
\label{eq67}
\d S' = \bd S' = 0
\ee
associated to the $\lt$ theory,
where $\xi$ and $\bx$ are the infinitesimal parameters of the
transformation. The $\a,\ba$ transformations in (\ref{eq57})
and (\ref{eq62}) can be related to the usual \cite{r7} 
${\rm Diff}\,S^2$ transformations of $v$ and $\bar{v}$ 
by
\bss\bea
\xi=\frac{1+\a}{2} \e
& \leftrightarrow  & \d v = \dot{\e} + \e \difflr_{\s} v
\label{eq61}
\\{}
\bx=\frac{1+\ba}{2} \bbe & 
\leftrightarrow &  \bd \bv = \dot{\bbe} + \bv \difflr_{\s} \bbe .
\label{eq66}
\eea\ess
It will also be useful to have the tangent-space form of these 
invariances
\bss
\bea
\label{eq68}
\d x^i & = & \xi^a e_a{}^i , \qquad \xi^a = 2 \xi \lt^a{}_b B^b
\\{}
\label{eq69}
\d B^a & = & \diffp \xi^a - \xi^b B^c (\hat{\omega}^-)_{bc}{}^a  + 
 \xi^b (\diffp x)^c (\hat{\omega}^+)_{cb}{}^a
\\{}
\label{eq70}
\d\bb^a & = & - \xi^b \bb^c (\hat{\omega}^-)_{bc}{}^a
\eea \label{u9} \ess
\bss
\bea
\label{eq71}
\bd x^i & = & \bx^a e_a{}^i , \qquad \bx^a = 2 \bx \lbt^a{}_b \bb^b
\\{}
\label{eq72}
\bd B^a & = & 
 - \bx^b B^c (\hat{\omega}^+)_{bc}{}^a
\\{}
\label{eq73}
\bd\bb^a & = &\diffm \bx^a - \bx^b \bb^c (\hat{\omega}^+)_{bc}{}^a  + 
 \bx^b (\diffm x)^c (\hat{\omega}^-)_{cb}{}^a
\eea\ess
where $(\diff_{\pm} x)^a= \diff_{\pm} x^i e_i{}^a$.
The semiclassical central charges of the spin-two gauged
sigma models are $(c_L,c_R)=({\rm rank}\, L,{\rm rank}\, \lt)$.

The original sigma model action $S_G$ is contained in the action $S'$ in two
independent ways:
\bea
{\rm (1)} & & L^{ab}=\lb^{ab}=L_G^{ab}, \qquad \lt=\lbt=0
\label{eq56}
\\{}
{\rm (2)} & & v=\bar{v}=1, \,\,\, \a=\bar{\a}=0, \quad \tilde{h}^{mn}=
\left( \begin{array}{cc} -1 & 0 \\ 0 & 1 \end{array} \right)
\label{eq55}
\eea
after trivially integrating out $B$ and $\bb$. The first is the choice to return to
$S_G$, while the second is the conformal gauge of the new
spin-two gauged sigma models. The Polyakov-gauged general non-linear
sigma model is also contained as the special case $L=\lb=0$,
$\lt=\lbt=L_G$ of these spin-two gauged sigma models (see Section~3.3).

\subsection{Non-Linear Form of the Spin 2 Gauged Sigma Models}

To go to the non-linear form of the action, we must integrate out the
auxiliary fields $B,\bb$ of $S'$ using the $B,\bb$ equations of motion
%
\bss
\bea
2 \a \lt^i{}_j B^j - \bb^i + \diffm x^i & = & 0
\label{75}
\\{}
2 \ba \lbt^i{}_j \bb^j - B^i + \diffp x^i & = & 0  .
\label{76}
\eea\ess
These can be solved for $B,\bb$ as
\bss
\bea
B & =& (1-4 \a\ba \lbt\lt)^{-1} (\diffp x + 2 \ba \lbt \diffm x)
\label{eq77}
\\{}
\bb & = & (1- 4 \a\ba\lt\lbt)^{-1} (\diffm x + 2\a \lt \diffp x)
\label{eq78}
\eea\ess
and substitution into $S'$ gives the non-linear form of the 
new spin 2 gauged sigma models
\bss
\bea
S & = & \int d^2 \xi \cl 
\label{n17} \\{}
\cl & = &  \ppe [( -G_{ij} + B_{ij} ) \diffp x^i \diffm x^j + m^T Nm]
\label{eq79}
\eea
\ess
where
\be
m=\left(
\begin{array}{c} \diffp x \\ \diffm x \end{array}
\right), \qquad
N = \left( 
\begin{array}{cc}
2 \a\lt(1-4 \a\ba\lbt \lt)^{-1} & (1-4 \a\ba \lt\lbt)^{-1} \\
(1-4 \a\ba\lbt\lt)^{-1} & 2 \ba\lbt (1-4 \a\ba\lt\lbt)^{-1}
\end{array}
\right)  .
\label{eq80}
\ee
The ${\rm Diff}\, S^2$ or spin-two gauge invariance of the non-linear
action
\bea
\d x & = & 2 \xi \lt (1-4\a\ba\lbt\lt)^{-1} (\diffp x + 2\ba \lbt \diffm x) 
\nonu 
\\{}
\bd x & = & 2 \bx \lbt (1-4\a\ba\lt\lbt)^{-1} (\diffm x + 2\a \lt \diffp x) 
\label{eq81}
\eea
(together with (\ref{eq57}) and (\ref{eq62})
is obtained by substitution of (\ref{eq77}),
(\ref{eq78}) in (\ref{eq58}) and (\ref{eq63}).
As in \cite{r7}, these nonlinear realizations of ${\rm Diff}\,S^2$ 
are generically new (see also (\ref{gurbe1})-(\ref{gurbe2})).

As in the linearized version, the original sigma model action $S_G$ 
is contained in the non-linear action $S$ in the same two
independent ways (\ref{eq55}), including the conformal
gauge choice $\a=\ba=0$. In this connection, we
note that the traceless covariantly-conserved symmetric 
stress tensor 
\be \label{n18}
\Theta^{mn} = \frac{2}{\sqrt{-\tilde{h}}} \frac{\delta S}{\delta \tilde{h}^{mn}}
\ee
reduces in the conformal gauge to 
\bea
& \Theta_{00} = \Theta_{11} = \tt_{++} + \tt_{--}
& \label{n19} \\{}
& \Theta_{01} = \Theta_{10} = \tt_{++} - \tt_{--}.
& \label{n20}
\eea
This identifies the $K$-conjugate metric $\tilde{h}_{mn}$ as the world-sheet 
metric of the $\lt$ theory, as expected from the generic affine-Virasoro action 
\cite{r7,r8}. 

Using the conformal gauge, we can also check directly that, in the
new actions, we have correctly gauged the $K$-conjugate part 
($\tt_{\pm\pm}$) of the spin-two symmetries (\ref{eq29}) of the
original sigma model $S_G$. According to (\ref{eq57}) and
(\ref{eq62}), the conformal gauge has a residual 
symmetry which leaves $\a=\ba=0$,
\bss
\bea
\diffm \xi&  = & \diffp \bx = 0 
\label{eq82}
\\{}
\d x^i & = & 2 \xi(\t^+) \lt^i{}_j \diffp x^j \equiv \tilde{\delta} x^i
\label{eq83}
\\{}
\bd x^i & = & 2 \bx(\t^-) \lbt^i{}_j \diffm x^j \equiv \tilde{\bar{\delta}} x^i
\label{eq84}
\eea\ess
and (\ref{eq83}) and (\ref{eq84}) are identical to our earlier
results in (\ref{eq29}).

We finally note that the Polyakov-gauged general non-linear sigma
model 
\be \label{gurbe1}
S_{GP}=
\frac{1}{8\pi \alpha'} 
\int d^2 \xi (-\sqrt{-\tilde{h}} \tilde{h}^{mn} 
 G_{ij} \partial_m x^i \partial_n x^j + 
 B_{ij} \partial_+ x^i \partial_- x^j)
\ee
(see (\ref{n16}))
is included as the special case 
\be
L=\bar{L}=0,\qquad \lt=\lbt=L_G
\ee
in this set of spin-two gauged sigma models. 
As in \cite{r7}, this is the only case where the
spin-two gauge transformations (\ref{eq81})
\bss
\bea
(\delta+\bar{\delta}) x & = & \epsilon \partial_+ x + 
\bar{\epsilon} \partial_- x \\{}
\left(
\begin{array}{c} \epsilon \\ \bar{\epsilon} \end{array} 
\right)
& = & 
\frac{1}{1-\alpha\bar{\alpha}} 
\left(
\begin{array}{cc} 1 & \alpha \\ \bar{\alpha} & 1 \end{array} \right)
\left(
\begin{array}{c} \xi \\ \bar{\xi} \end{array} 
\right)
\eea \label{gurbe2}
\ess
are ordinary diffeomorphisms and the K-conjugate metric $\tilde{h}^{mn}$
is the world-sheet metric of the theory.

\section{Example: The Spin-Two Gauged WZW Actions}

In this section we study the spin-two gauged WZW actions as an
explicit example of the spin-two gauged sigma models
(\ref{e39}) and (\ref{qqqq}), and identify this set of actions as
the generic affine-Virasoro action \cite{r7,r8,r9}. 

To begin we choose the background sigma model $S_G$ to
be the sigma model form of the
WZW action $S_{WZW}$ on simple Lie $g$, whose explicit form is given
in Appendix~A. The sigma model data for the WZW action is \cite{r14}
%
%
\bss
\bea
& y^i=\frac{1}{\sqrt{\a'}} x^i, \qquad i=1,\ldots,\dim\, g
& \label{eq86}
\\{} & G_{ab} = k \eta_{ab}, \qquad a,b=1,\ldots,\dim\, g & 
\label{eq87a} \\{}
& e_i=-ig^{-1} \frac{\partial}{\partial y^i}g=e_i{}^a T_a,\qquad
 \bar{e}_i=-ig \frac{\partial}{\partial y^i}g^{-1}=\bar{e}_i{}^a T_a
& \label{eq87b} \\{} &
\tr(T_a T_b) = y G_{ab} & \label{eq87c} \\{}
& \hat{\omega}^+_{ab}{}^c=0, \qquad \hat{\omega}^-_{ab}{}^c = 
-\frac{1}{\sqrt{\a'}} f_{ab}{}^c .  &
\label{eq87}
\eea
\label{u1}
\ess
Here $y^i$ are the dimensionless coordinates of the WZW action, 
$g \in G$ is the group element,
$\eta_{ab}$ and
$f_{ab}{}^c$ are respectively the Killing metric and structure constants of  $g$
and $k$ is the (high) level of affine $g$. The asymmetry of the spin connections
$\hat{\omega}^{\pm}$ in (\ref{eq87})
follows because we have identified the sigma model
vielbein with the left-invariant vielbein $e_i{}^a$ on the group manifold.

In this case, the stress tensors 
$T^G_{\pm\pm}$ of the background sigma model are the (high-level
forms of) the affine-Sugawara constructions $T^g_{\pm\pm}$
on $g$, and the $K$-conjugation relations (\ref{n11}) read
\bss\bea
& T^g_{\pm\pm}= T_{\pm\pm} + \tt_{\pm\pm} &
  \label{n22a} \\{}
&  K_g: \qquad L^{ab}_g = L^{ab} + \lt^{ab} =
\lb^{ab} + \lbt^{ab} & 
\label{n22b} \\{}
& L_g^{ab} = \frac{G^{ab}}{2} = \frac{\eta^{ab}}{2k} & 
\label{n22c}
\eea \label{n22} \ess
in this case. With $L_g^{ab}=\eta^{ab}/(2k + Q_g)$ and the
exact solutions $L,\lt$ of the Virasoro master equation, these
relations are exact at the quantum level in the general
affine-Virasoro construction. It follows that the corresponding
spin-two gauged form (\ref{e39}) of the WZW action must be
equivalent to the generic affine-Virasoro action,
which describes a very large class of irrational conformal
field theories. As we shall see, 
this is not difficult to check.

To realize the spin-two gauged WZW action (\ref{e39}), we 
need the solutions of the covariant constancy
conditions (\ref{q1}) for the WZW background (\ref{u1}).
These were given in \cite{r14},
\bss\bea
& L^{ab}={\rm constant} & 
\label{n23}
\\{} &
 \lb'^{ab}=\lb^{cd} \Omega_c{}^a \Omega_d{}^b = {\rm constant}
& \label{eq88}
\\{} &
\frac{\diff}{\diff y^i} \Omega_{a}{}^b + e_i{}^d f_{da}{}^c \Omega_c{}^b =0
& \label{eq89}
\eea \label{u3} \ess
where $\Omega$ is the adjoint action of $g$. The $K$-conjugate form 
of these solutions
\bss\bea
\lt^{ab} & = & {\rm constant} 
\label{n24} \\{}
\lbt'^{ab} & = & \lbt^{cd} \Omega_c{}^a \Omega_d{}^b={\rm constant}
\label{n25}
\eea \label{u4} \ess
then follows from eq (\ref{n22b}). The following 
associated definitions
\bss\bea
\label{eq90}
& B'_a = \frac{1}{2\sqrt{\a'}} B_a , \qquad \bb'_a = -\frac{1}{2\sqrt{\a'}}
(\Omega^{-1})_a{}^b \bb_b
& \\{}
\label{eq91}
& B'=B'^a T_a, \qquad \bb'=\bb'^a T_a
& \\{} & 
\xi'=2\xi , \qquad  \bx'=2\bx
\label{eq92}
& \\{} & 
\diff=\frac{1}{2} \diffp, \qquad \diffb = \frac{1}{2} \diffm
\label{eq93}
& \eea \label{u5} \ess
will also be useful.

With the WZW results (\ref{u1}), (\ref{u3}) and (\ref{u4}) and the
definitions (\ref{u5}), we find that the spin-two
gauged WZW action (\ref{e39}) can be written as
\bss
\bea
S'_{AV} & = & S_{WZW} + \int d^2 \xi \, \Delta \cl_B
\label{eq94}
\\{}
\Delta \cl_B & = & \frac{\a}{\pi y^2} \lt^{ab} \tr(T_a B') \tr(T_b B')
\nonu
\\{}
& & +  \frac{\ba}{\pi y^2} \lbt'^{ab} \tr(T_a \bb') \tr(T_b \bb')
\nonu \\{}
&& -\frac{1}{\pi y} \tr(\bar{D} g D g^{-1})
\label{eq95}
\\{}
D& =& \diff + iB' , \qquad \bar{D} = \diffb + i\bb'
\label{eq96}
\\{} \lt^{ab} & = & 2 \lt^{ac} G_{cd} \lt^{db}, \qquad
\lbt^{ab}  =  2 \lbt^{ac} G_{cd} \lbt^{db}
\label{eq96a}
\eea \label{u25} \ess
where $g\in G$ is the group element of the WZW action $S_{WZW}$.
Moreover, the tangent-space form (\ref{u9}) of the ${\rm Diff}\, S^2$
invariance of the spin-two gauged WZW action can be put in the form
\bss\bea
&  \d\a = -\diffb \xi' + \xi' \difflr \a, \qquad
\d\ba = - \diff \bx' + \bx' \difflr \ba & 
\label{eq97}
\\{}
 & \d g = gi\lambda - i\bar{\lambda} g &
\label{eq98}
\\{}
 & \d B'=\diff \lambda + i[B,\lambda], \qquad \d \bb = 
\diffb \bar{\lambda} + i[\bb,\bar{\lambda}] &
\label{eq99}
\\{}
 & \lambda=\lambda^a T_a, \qquad \bar{\lambda} = \bar{\lambda}^a T_a &
\label{eq100}
\\{}
& \lambda^a = 2 \xi' \lt^{ab} B_b', \qquad
\bar{\lambda}^a = 2 \bx' \lbt'^{ab} B_b'  . & 
\label{eq101}
\eea \label{e47} \ess
If we also make the special choice\footnote{In the
language of \cite{r7}, the subscript $\infty$ means
the high-level limit $L^{ab}_{\infty}=P^{ab}/2k$ of
any high-level smooth solution of the Virasoro master 
equation, in agreement with (\ref{eq96a}). More generally,
the classical inverse inertia tensors $L,\lt,\bar{L},\lbt$
of the spin-two gauged sigma models are the
semiclassical limit of the one-loop 
inverse inertia
tensors of \cite{r14}}
\bss
\bea
& \lb'^{ab} = L^{ab} \equiv L_{\infty}^{ab} & 
\nonu \\{}
& \lbt'^{ab} = \lt^{ab} \equiv \lt_{\infty}^{ab} & 
\label{n29}
\eea\ess
we see that (\ref{u25}) and (\ref{e47}) are precisely the generic
affine-Virasoro action \cite{r7,r8,r9} 
(in the form given in \cite{r9})
and its spin-two
gauge invariance. Without the choice 
(\ref{n29}), the
spin-two gauged WZW actions 
(\ref{u25}) offer a mild
generalization of the affine-Virasoro action, which may
describe non-diagonal constructions of the
corresponding irrational conformal field theories.

We emphasize that the generic affine-Virasoro action describes only those
generic constructions in the Virasoro master equation which have no larger
local symmetry than the ${\rm Diff}\,S^2$ associated to the $K$-conjugate
theory. In particular, the generic action does not describe the local Lie
$h$-invariant CFTs \cite{r16}. This is the large, generically irrational
set of all CFTs with an additional local Lie $h$ invariance, including
the coset constructions as the simplest case. The actions for the local 
Lie $h$-invariant CFTs will be obtained in the following section.

\section{Example: The Spin-Two Gauged Coset Constructions}

In this section, we study another special case of the spin-two
gauged sigma models, namely the spin-two
gauged $g/h$ coset constructions. This class of
examples includes the spin-two
gauged WZW actions in the formal limit when $g\supset h\ra 0$.

In this case, the background sigma model $S_G$ is the sigma
model description $S_{g/h}$ of
the coset constructions, that is, the sigma model form of the
spin-one gauged WZW actions (see Appendix~A).
The $K$-conjugation relations read
\bss\bea
&K_{g/h} : \qquad T_{\pm\pm}^{g/h} = T_{\pm\pm} + \tt_{\pm\pm} & 
\label{n30} \\{}
& L_{g/h}^{ab} = L_g^{ab}- L_h^{ab} = 
L^{ab} + \lt^{ab} = \lb^{ab} + \lbt^{ab}
\label{n31} 
\eea\ess
in this case, where $T^{g/h}_{\pm\pm}$ are the (high-level forms of
the) stress tensors of the coset constructions. $K$-conjugation
through the coset constructions, as seen in (\ref{n30}), is known to
be exact at the quantum level for the local Lie $h$-invariant CFTs
\cite{r16}
of the Virasoro master equation, so we expect and will find that our
action (\ref{e39}) for the general spin-two gauged coset
construction is the action for this large class of
generically irrational conformal field theories.

\subsection{Inverse Inertia Tensors in the Coset Backgrounds}

To realize the spin-two gauged coset actions (\ref{e39}) in this case,
our first task is to solve the covariant-constancy conditions
(\ref{q1}) in the background of the general $g/h$ coset 
construction. For this, we will need the following data,
derived in Appendix~A, for the sigma model description
of the coset construction,
%
%
%
\bss
\bea
& y^i = \frac{1}{\sqrt{\a'}} x^i &
\label{eq104}
\\{}
& G_{ij} = e_i{}^a G_{ab} e_j{}^b = \bar{e}_i{}^a G_{ab} \bar{e}_j{}^b &
\label{eq105}
\\{}
 & e_i{}^a = (1+M)^a{}_{\mu} L_i{}^{\mu} &
\label{eq106}
\\{}
\label{eq107}
& \bar{e}_i{}^a = -e_i{}^b \Lambda_b{}^a = -L_{i}{}^{\mu}  
((1+M)\Omega)_{\mu}{}^a &
\\{}
& \Lambda=P_{g/h} (1+M)\Omega P_{g/h} ,\qquad \Lambda \Lambda^T = P_{g/h} &
\label{eq108}
\\{}
& M=\Omega P_h N P_h , \qquad N=(1-P_h \Omega P_h)^{-1} & 
\label{eq109}
\\{}
 &  (\hat{\omega}_i^+)_a{}^b = \frac{1}{\sqrt{\a'}} L_i{}^{\mu} M_{\mu}{}^A f_{Aa}{}^b & 
\label{eq111} 
\\{}
&  (\hat{\omega}_i^-)_a{}^b = \Lambda_a{}^c \hat{\omega}^-(\Lambda)_c{}^d
(\Lambda^{-1})_d{}^b + \partial_i \Lambda_a{}^c (\Lambda^{-1})_c{}^b & 
\label{eq112}
\\{}
& \hat{\omega}^-(\Lambda)_a{}^b = -
\frac{1}{\sqrt{\a'}} N^A{}_B L_i^B f_{Aa}{}^b & 
\label{eq113}
\\{}
& i,j  =  1,\ldots, \dim \, g/h \quad ({\rm curved}) & 
\nonu
\\{}
& a,b  =  1,\ldots, \dim \, g/h \quad ({\rm flat}) & 
\nonu
\\{}
& \mu,\nu  =  1,\ldots, \dim \, g \quad ({\rm flat}) & 
\nonu
\\{}
& A,B  =  1,\ldots, \dim \, h \quad ({\rm flat})   . & 
\label{eq110}
\eea \label{u11} \ess
Here $P_h$ and $P_{g/h}$ are projectors onto $h$ and $g/h$,
$\Omega$ is the adjoint action of $g$, and $L_i{}^{\mu}$ is the
restriction to $i=1,\ldots,\dim\,g/h$ of the left invariant 
vielbein $L$ on the group manifold.

The quantities $\Lambda$, $e$ and $\bar{e}$ 
in (\ref{u11}) are the analogues on the
coset of the adjoint action 
$\Omega$
and the left and right invariant vielbeins $L$ and
$R$ on
the group manifold (called $e$ and $\bar{e}$ in Section~4).
Indeed, as a check, we note the agreement with the WZW data
(\ref{u1}),
\bea
& & \Lambda \ra\Omega, \qquad 
e_i{}^a \ra L_i^a, \qquad \bar{e}_i^a \ra R_i^a=-L_i{}^{\mu} 
\Omega_{\mu}^a 
\nonu \\{} & & 
 (\hat{\omega}^+_a)_b{}^c \ra 0 , \qquad (\hat{\omega}^-_a)_b{}^c \ra 
- \frac{1}{\sqrt{\a'}} f_{ab}{}^c 
\label{eq114}
\eea
in the formal limit $h\rightarrow 0$, $g/h \ra g$.

As in the case of the WZW background, $\hat{\omega}^-$
is more complicated than $\hat{\omega}^+$, but
$\hat{\omega}^-$ is a gauge transformation of $\hat{\omega}^-(\Lambda)$,
whose form is similar to $\hat{\omega}^+$. We may bring both
covariant-constancy conditions into  similar form
%
%
%
%
%
%
\bss
\bea
\label{eq116}
& \delhp_i(\hat{\omega}^+)  L^{ab}=
\delhm_i (\hat{\omega}^-(\Lambda)) \lb'^{ab}=0
& \\{}
& 
\label{eq117}
\lb'^{ab} \equiv  \lb^{cd} \Lambda_c{}^a \Lambda_d{}^b
& \eea\ess
%
by the definition in (\ref{eq117}). In (\ref{eq116}) we have
explicitly indicated the spin connections in the
gradients $\delh^{\pm}_i$.

In both problems (\ref{eq116}) the relevant spin connections 
$\hat{\omega}^{\pm}\sim\hat{\omega}$ are proportional to
the structure constants, and we take them in the block-diagonal
form
\bss
\bea
& \delh_i(\hat{\omega}) L^{ab}=0 & 
\label{n40} \\{}
\label{eq119} &
(\hat{\omega}_i)_a{}^b = 
R_i{}^A f_{Aa}{}^b = 
\left(
\begin{array}{ccc|c}
B^1_i & & & \\
& \ddots & & 0 \\
& & B^r_i & \\
\hline
& 0 & & 0
\end{array}
\right) & 
\\{}
\label{eqn41} &
L_a{}^b = 2L_a{}^c L_c{}^b & 
\eea \label{e55} \ess
where the blocks $B_i^s$ are irreps of $h\subset g$.
The general solution to the system (\ref{e55}) has
been discussed in \cite{r15}. The result is 
\bss\bea
\label{eq120}
& L_a{}^b =\frac{1}{2} \left(
\begin{array}{cccc|c}
{\bf 1}_1 \theta_1 & & & & \\
& {\bf 1}_2 \theta_2 & & & \\
& & \ddots & & 0 \\
& & & {\bf 1}_r \theta_r & \\
\hline
& & 0 & & \theta
\end{array}
\right) & 
\\{}
\label{eq121}
& \theta_i=0,1; \qquad \theta^2 = \theta
& \eea
\label{u15}
\ess
where ${\bf  1}_s$ are unit matrices and the projectors
$\theta_r,\theta$ are constants, so that (\ref{u15}) solves
(\ref{e55}) in the form
\be
\label{eq122}
\partial_i L^{ab} = L^{c(a} f_{cA}{}^{b)}=0 .
\ee
Here $(ab)$ means symmetrization with respect to
the indices $a$ and $b$.

In summary, we have shown that a large class of inverse inertia
tensors exist in the coset backgrounds:
\bss\bea
\label{n42}
& L^{ab} = {\rm constant} & 
\\{}
\label{n43}
& \lb'^{ab} = \bar{L}^{cd} \Lambda_c{}^a \Lambda_d{}^b = 
{\rm constant} &
\\{}
\label{n44}
& L^{ab} = 2 L^{ac} G_{cd} L^{db} , \qquad
\lb'^{ab} = 2 \lb'^{ca} G_{cd} \lb'^{db} & 
\\{}
\label{n45} &
L^{c(a} f_{cA}{}^{b)} = \lb'^{c(a} f_{cA}{}^{b)} = 0 &
\eea \label{u17}\ess
where the constant solutions of (\ref{n44}), (\ref{n45}) have the
form (\ref{u15}). The same results with $L\ra\lt$, $\lb\ra\lbt$ follow for the
$K$-conjugate inverse inertia tensors. The spin-two gauged coset actions
(\ref{e39}) are realized for any solution of (\ref{u17}).

\subsection{Identification with the Local Lie $h$-Invariant CFTs}


We now turn to the identification of the new spin-two gauged
coset actions as the actions of the local Lie $h$-invariant CFTs
\cite{r16}. 

The Lie $h$-invariant CFTs
 are found in the Virasoro master equation on $g\supset h$, where the 
general inverse inertia tensor $L$ is labelled as 
\bss
\bea
\label{n46}
& L^{\mu\nu} = {\rm constant},\qquad \mu=(A,a),\quad \nu=(B,b) &
\\{}
\label{n47}
& L^{\mu\nu} = 2L^{\mu\rho} G_{\rho\sigma} L^{\sigma\nu} 
+ {\cal O}(k^{-2}) & 
\eea\ess
in the present notation (see (\ref{u11})).
The relation in (\ref{n47}) is the semiclassical or 
high-level form of the 
Virasoro master equation on $g$, which is all we will
need for this discussion (see also Appendix~B).
 The necessary and sufficient condition that the
CFT is Lie $h$-invariant is that the inverse inertia tensor is invariant
under infinitesimal $h$ transformations 
\be
\label{eq123}
L^{\rho(\mu} f_{\rho A}{}^{\nu)}=0, \qquad A=1,\ldots,\dim\, h .
\ee 
The distinction between global and local Lie $h$-invariant
theories is that the $h$-currents $J_A$ are respectively $(1,0)$ and
$(0,0)$ operators of $T$
\bea
\label{eq163}
 {\rm global:}&  & [T(m),J_A(n)]=-n J_A(m+n) 
\\{}
\label{eq164}
 {\rm local:}& &  [T(m),J_A(n)]=0 
\eea
when $T=L^{\mu\nu}:J^+_{\mu} J^+_{\nu}:$ is a Lie $h$-invariant 
construction. 
The coset constructions are the simplest examples of local 
Lie-$h$-invariant CFTs, but both classes are vast, with generically
irrational central charge.

The local Lie $h$-invariant CFTs always occur in $K$-conjugate pairs
\be
K_{g/h}: \quad T_{g/h}=T+\tt
\ee
where the conjugation is through the coset constructions.

Eq. (\ref{eq164}) also tells us that no $h$-currents are present
in the stress tensor of a local Lie $h$-invariant CFT at high level
\be
\label{eq124}
L^{ab} = {\cal O}(k^{-1}) , \qquad L^{AB}={\cal O}(k^{-2}),
\qquad L^{Aa}={\cal O}(k^{-2}) 
\ee  
and hence that the coset-valued part of $L$ is semiclassically 
dominant for the local Lie $h$-invariant CFTs.
In this case, (\ref{n47}) and (\ref{eq123}) reduce to 
\bss\bea
\label{n50}
& L^{ab} = 2L^{ac} G_{cd} L^{ab} + {\cal O}(k^{-2}) & 
\\{}
\label{n51} &
L^{c(a} f_{cA}{}^{b)} =
{\cal O}(k^{-2}) &
\eea \label{u19} \ess
and the same equations hold for the $K$-conjugate theory
$L\ra\lt$ because both $T$ and $\tt=T_{g/h}-T$ are
local Lie $h$-invariant CFTs under 
conjugation through $g/h$.

The semiclassical conditions (\ref{u19}) are
the same conditions (\ref{n44}), (\ref{n45})
we found for the allowed classical 
inverse inertia tensors in the
coset backgrounds, and this completes the 
identification of our spin-two gauged coset actions
as the actions of the local Lie $h$-invariant
CFTs.

\subsection{Spin 2 and Spin 1 Gauged Form of the New Actions}

In this section, we find spin~2 and spin~1 gauged 
actions which are equivalent descriptions of the
spin~2 gauged coset constructions/local Lie $h$-invariant
CFTs of the previous section.

Although the generic affine-Virasoro action (\ref{u25}) is not applicable
to the local Lie $h$-invariant CFTs, our new spin~2 gauged coset
actions should be equivalent to a spin~1 gauging 
(by $h\subset g$)
of the affine-Virasoro
action for local Lie $h$-invariant CFTs:
%
%
%
%
\bss
\bea
\label{eq126}
S_{2,1} & =& S'_{AV}|_{{\rm local}\,\,{\rm Lie}\,h} + 
\frac{1}{\pi y} 
\int d^2 \xi \,
\tr(A\bar{A}-A\bb'-\bar{A}B')
\\{}
\label{eq127}
S'_{AV}|_{{\rm local}\,\,{\rm Lie}\,h} 
& =&  S_{WZW} + \int d^2 \xi [ 
-\frac{1}{\pi y} \tr(\bar{D} g D g^{-1}) \nonu
\\{} && \qquad \qquad \qquad \quad
+\frac{\alpha}{\pi y^2} \lt^{ab}
\tr(T_a B') \tr(T_b B') 
\nonu \\{} && \qquad \qquad \qquad \quad
+\frac{\ba}{\pi y^2} \lbt'^{ab} \tr(T_a\bb') \tr(T_b \bb')]
\eea \label{u31} 
\bea
\label{nn42}
& \lt^{ab} = {\rm constant} & 
\\{}
\label{nn43}
& \lbt'^{ab} = \lbt^{cd} \Lambda_c{}^a \Lambda_d{}^b = 
{\rm constant} &
\\{}
\label{nn44}
& \lt^{ab} = 2 \lt^{ac} G_{cd} \lt^{db} , \qquad
\lbt'^{ab} = 2 \lbt'^{ca} G_{cd} \lbt'^{db} & 
\\{}
\label{nn45} &
\lt^{c(a} f_{cA}{}^{b)} = \lbt'{}^{c(a} f_{cA}{}^{b)} = 0  . &
\eea \label{u21} \ess
Here $S'_{AV}|_{{\rm local}\,\,{\rm Lie}\,h}$ is nothing
but the generic affine-Virasoro action $S'_{AV}$ on $g$ in 
(\ref{u25}), now evaluated for any 
$\lt,\lbt'$ which are local Lie $h$-invariant, as shown in
(\ref{nn45}). The indices 
$a,b$ in (\ref{u21}) take the values $1,\ldots,\dim\, g/h$,  
because $\lt,\lbt'$ are dominated by their values
on the coset, and the auxiliary fields
$B^i,\bar{B}^i$ are $g$-valued.
On the other hand, the spin~1 gauge fields
$A,\bar{A}$ are valued on $h$. 
This action has the 
following spin~2 and spin~1 gauge symmetries
\bss\bea
&  \d\a = -\diffb \xi' + \xi' \difflr \a, \qquad
\d\ba = - \diff \bx' + \bx' \difflr \ba & 
\label{eqq97}
\\{}
 & \d g = gi\lambda - i\bar{\lambda} g &
\label{eqq98}
\\{}
 & \d B'=\diff \lambda + i[B',\lambda], \qquad \d \bb' = 
\diffb \bar{\lambda} + i[\bb',\bar{\lambda}] &
\label{eqq99}
\\{}
& \d A=\diff \epsilon_H + i[A,\epsilon_H], \qquad \d \bar{A} = 
\diffb \epsilon_H + i[\bar{A},\epsilon_H] &
\label{eqq99a}
\\{}
 & \lambda=\lambda^a T_a+\epsilon_H, \qquad \bar{\lambda} = \bar{\lambda}^a T_a +\epsilon_H
&
\label{eqq100}
\\{}
& \lambda^a = 2 \xi' \lt^{ab} B_b', \qquad
\bar{\lambda}^a = 2 \bx' \lbt'^{ab} B_b'  & 
\label{eqq101}
\eea
\label{e516}
\ess
where $\epsilon_H$, which is is $h$-valued, is the parameter for the spin 1 symmetry, and
$\xi'$, $\bx'$ are the parameters for the spin 2 symmetry. 

From the viewpoint of stress tensors in the Virasoro master equation
on $g$, the action (\ref{u21}) describes the local Lie $h$-invariant
stress tensor $T_{llh}=T(L)$,
which appears in two $K$-conjugation relations \cite{r16}
\bss\bea
T_g & = & T_{llh} + T_{glh} \\{}
T_{g/h} & = & T_{llh} + \tilde{T}_{llh}
\eea\ess
where $\tilde{T}_{llh}=T(\tilde{L})$ is also local Lie $h$-invariant and 
$T_{glh}$ is a global Lie $h$-invariant theory. Combining these
relations, we find that
\bss\bea
\tilde{T}_{llh} & = & T_{glh}-T_h \\{}
[T_{llh},\tilde{T}_{llh}] & = & [T_{llh},J_A] \,\,\, = \,\,\,
[\tilde{T}_{llh},J_A] \,\,\,  = \,\,\, 0
\eea\ess
where $A$ labels the $h$-currents. The commuting operators
$\tilde{T}_{llh}$ and $J_A$ generate the local spin-two and
spin-one symmetries by which the theory $T_{llh}=T(L)$
is gauged in (\ref{u21}).
 For the special case of the cosets themselves ($T_{llh} =
T_{g/h}$), we see that $\tilde{T}_{llh}= \tilde{L} =0$ and the spin-two gauging
decouples in (\ref{u21}), leaving the ordinary spin-one gauged WZW action 
\cite{r22,r23,kahs,kas}.

As a check, we integrate out $A$ and $\bar{A}$ in (\ref{u21}) to
obtain
\bss
\bea
\label{eq128}
S' & =& S_{WZW} + \int d^2 \xi \, \Delta \cl'_B
\\{}
\label{eq129}
\Delta \cl'_B & = & \frac{\alpha}{\pi y^2} \lt^{ab}
\tr(T_a B') \tr(T_b B') 
\nonu \\{}
& & +\frac{\ba}{\pi y^2} \lbt'^{ab} \tr(T_a\bb') \tr(T_b \bb')
\nonu \\{}
& & -\frac{1}{\pi y} \tr(( \diffb + i \bb') g (\diff+i B') g^{-1} )
\nonu \\{}
& & -\frac{1}{\pi} B_A' G^{AB} \bb'_B
\eea \label{e518} \ess
where $\Delta \cl'_B$ differs from $\Delta \cl_B$ in
(\ref{eq94}) only by the last term in (\ref{e517}).
This action still enjoys the spin~2 and spin~1 symmetries
given in (\ref{e516}) except for
(\ref{eqq99a}).
Finally, 
we have shown by gauge fixing the spin~1 symmetry and
integrating out $B_A',\bb_A'$  that we get exactly the
sigma model form (\ref{e39}) of the spin~2 gauged
coset constructions, with the identifications
\bss\bea
\label{eq130}
& y^i = \frac{1}{\sqrt{\a'}} x^i  &
\\{}
\label{eq131}
& B'_a = \frac{1}{2\sqrt{\a'}} B_a &
\\{}
\label{eq132}
& \bb'_a = -\frac{1}{2\sqrt{\a'}} (\Lambda^{-1})_a{}^b \bb_b &
\\{}
\label{n60}
& \lt^{ab} = {\rm constant} & 
\\{}
\label{eq133}
& \lbt'^{ab} = \lbt^{cd} \Lambda_c{}^a \Lambda_d{}^b = {\rm constant}.&
\eea
\label{e517}
\ess
It follows that the spin~2 and spin~1 gauged actions (\ref{u31}) 
or (\ref{e518}) 
are equivalent to the spin~2 gauged coset
actions (\ref{e39}), and any of these are equivalent descriptions 
of the local Lie $h$-invariant CFTs.


\section{The Doubly-Gauged Actions}

As a final topic, we briefly discuss the doubly-gauged sigma 
model actions, which include the doubly-gauged WZW
actions of Refs. \cite{r7,r8,r9}. In these actions the background sigma
model is gauged by both the $L$ and the $\lt$ theories, resulting
in a ${\rm Diff}\,S^2 \times {\rm Diff}\,S^2$ symmetry with
two spin-two gauge fields.

Following the references, one begins with the doubly-gauged Hamiltonian
\bea
\label{eq134}
\ch_2 & = &  u \cdot T + v \cdot \tt \nonu \\{}
&  = & u T_{++} + \bar{u} T_{--} + v \tt_{++} + \bar{v} \tt_{--} 
\nonu \\{} & = & 
\ppe [  (u L^{ab} + v \lt^{ab}) J_a^+ J_b^+ \nonu \\{}
& & \qquad
 +(\bar{u} \lb^{ab} + \bar{v} \lbt^{ab}) J_a^- J_b^- ]
\eea
which exhibits a $({\rm Diff}\,S^1)^4$ symmetry generated by
all four stress tensors $T_{\pm\pm}$ and $\tt_{\pm\pm}$. The extra
multipliers $u$ and $\bar{u}$ form a second spin-two gauge field 
$h_{mn}$ of the form (\ref{n16}) with $v,\bar{v}\ra u,\bar{u}$
which can be identified, as in (\ref{n18}), (\ref{n19})
 as the world-sheet metric of the $L$ theory.

Adding the usual auxiliary fields $B$ and $\bb$, one finds the
doubly-gauged sigma models
\bss\bea
\label{eq135}
\cl_2 & =&  \cl_G + \ppf [  (
\a \lt^{ij} + \beta L^{ij}) B^i B^j \nonu \\{}
& & \qquad \qquad \quad + (\ba \lbt^{ij} + \bar{\beta} \lb^{ij})
\bb^i \bb^j \nonu \\{}
& & \qquad \qquad \quad
-(B^i-\diffp x^i) G_{ij} (\bb^j - \diffm x^j) ]
\eea
\be
\label{eq136}
\a=\frac{1-v}{1+v}, \quad
\ba=\frac{1-\bar{v}}{1+\bar{v}}, \quad
\beta=\frac{1-u}{1+u}, \quad
\bar{\beta}= \frac{1-\bar{u}}{1+\bar{u}}
\ee \label{u41} \ess
where $\cl_G$ is the sigma model Lagrange density in (\ref{u51}).
The conditions (\ref{qqqq}) (and the equivalent conditions
(\ref{e210})) apply here as well.
In these actions, the $({\rm Diff}\,S^1)^4$ 
symmetry of the Hamiltonian (\ref{eq134}) 
is promoted to a ${\rm Diff}\,S^2\times {\rm Diff} \, S^2$
invariance
\be
\delta S_2  =  \bd S_2 =0
\label{eq137}
\ee
where the explicit form of the invariance is
\bss\bea
& \delta \a  =  -\diffm \xi + \xi \difflr_+ \a, \qquad
%
\delta \beta  =  -\diffm \zeta + \zeta \difflr_+ \beta  & \\{}
& \delta x^i =  \xi^i  =  2(\xi \lt^i{}_j + \zeta L^i{}_j)B^j &
\\{}
\label{eq139}
& \delta B^i  =  \diffp \xi^i - (B^j -\diffp x^j) \xi^k
 (\gh^+)_{jk}{}^i &
\\{}
\label{eq140}
& \delta \bb^i  =  -\xi^j \bb^k (\gh^-)_{jk}{}^i &
\eea\bea
& \bd \ba  =  -\diffp \bx + \bx \difflr_- \ba, \qquad
\bd \bar{\beta} = - \diffp \bar{\zeta} + 
\bar{\zeta} \difflr_- \bar{\beta} &
\label{eq141}
\\{}
& \bd x^i = \bx^i  = 
 2(\bx \lbt^i{}_j + \bar{\zeta} \bar{L}^i{}_j) \bb_j &
\label{eq142}
\\{}
& \bd B^i  =  -\bx^j B^k (\gh^+)_{jk}{}^i & 
\label{eq143}
\\{}
& \bd \bb^i  =  -\diffm \bx^i - (\bb^j - \diffm x^j)
\bx^k (\gh^-)_{jk}{}^i  . &
\label{eq144}
\eea\ess
With the discussion of Section~4, we see
that the doubly-gauged WZW action \cite{r7,r8,r9}
is included in (\ref{u41})
when the background sigma model is the WZW action. Moreover,
using the results of Section~5, we see that the
doubly-gauged coset
actions are included in (\ref{u41})
for each $K$-conjugate pair of local Lie
$h$-invariant CFTs. Using the two world-sheet metrics
$h_{mn}$ and $\tilde{h}_{mn}$ of the $K$-conjugate
theories, one application for such constructions is
discussed in \cite{r9}.

In the doubly gauged actions (\ref{u41}), 
the pair $L,\lt$ of $K$-conjugate 
CFTs are included symmetrically: To describe the CFT $L$,
one views $h_{mn}$ as a fixed world-sheet metric and integrates
out the $K$-conjugate metric $\tilde{h}_{mn}$, and vice-versa
to describe the CFT $\tilde{L}$.

An alternative procedure is to integrate out both spin-two
gauge fields $h_{mn}$ and $\tilde{h}_{mn}$, which defines
a new class of string theories where the physical states are
simultaneously primary under the $K$-conjugate pairs of commuting
Virasoro generators
($T_{++}$, $T_{--}$, $\tt_{++}$ and $\tt_{--}$). The first
example of this kind of string theory was the ``spin-orbit''
model of \cite{r6} (see also \cite{r3}) and this new class of string
theories may also be related to the models of \cite{r19}.
We note in particular that Virasoro biprimary fields \cite{rnew,r3}
have arisen naturally in both contexts.

\section{Conclusions and Discussion}

We have obtained the action formulation of a large class of
new CFTs whose stress tensors were recently 
constructed 
\cite{r14,r15} at the one-loop level
in the background of the general
conformal non-linear sigma model. The actions are generically new
spin-two gauged sigma models
\bss
\bea
\label{yeq52}
S'& =& \int d^2 \xi\, \cl'
\\{}
\cl' & = & \cl_G + \ppf [\alpha \lt_{ij} B^i B^j +
\bar{\alpha} \lbt_{ij} \bar{B}^i \bar{B}^j 
\nonu \\{}
& & \qquad \qquad
 \quad -(B^i-\diffp x^i) G_{ij} (\bar{B}^j - \diffm x^j) ]
\label{yeq54}
\eea \label{u7}
\bea
& \delhp_i \lt_j{}^k = \delhm_i \lbt_j{}^k =0 & 
\label{yq1}
\\{}
& \lt_i{}^j = 2 \lt_i{}^k \lt_k{}^j , \qquad
\lbt_i{}^j = 2 \lbt_i{}^k \lbt_k{}^j & 
\label{yq2}
\eea \label{e71}
\ess
where $\int d^2 \xi \cl_G$ is the action of the general non-linear
sigma model. The spin-two gauge symmetry of these actions is 
associated with $K$-conjugation $K_G$ through the background conformal
sigma models. The spin-two gauged sigma models contain (at least)
the following special cases

$\bullet$\quad The general non-linear sigma model

$\bullet$\quad The Polyakov-gauged general non-linear sigma model

$\bullet$\quad The spin-two gauged WZW actions

$\bullet$\quad The spin-two gauged $g/h$ coset actions

\noindent
which we have discussed in some detail:
The spin-two gauged WZW actions (associated to $K_{g}$ conjugation
through the affine-Sugawara construction) describe the generic CFT in the
Virasoro master equation, and these actions are nothing but
the sigma model form of the generic 
affine-Virasoro action \cite{r7,r8,r9}.
The spin-two gauged coset actions
(associated to $K_{g/h}$ conjugation through the coset constructions)
describe the local Lie $h$-invariant CFTs \cite{r16}. This is the set of
all CFTs in the Virasoro master equation with an extra $h$ gauge
symmetry, including the coset constructions as the simplest case.

Beyond these examples, new CFTs are obtained for every solution of
the conditions (\ref{yq1}), (\ref{yq2}) in the background of the
conformal sigma model. The necessary and sufficient conditions for
the solutions of these conditions is discussed in \cite{r14,r15}. 
It is expected
that the class of CFTs described by the spin-two gauged sigma
models is vast, one hint being the observation of many
other $K$-conjugation covariances \cite{r16} in the Virasoro
master equation, beyond $K_g$ and $K_{g/h}$ discussed here.
These include $K$-conjugation through $g+h$ and the 
general affine-Sugawara nests $g/h_1/\ldots/h_n$.
Beyond this, it is reasonable to
expect many new $K_G$-covariances not associated to group 
manifolds.

Another direction for generalization is as follows. The spin-two
gauged sigma models describe only generic CFTs whose local
symmetry is associated only to $K$-conjugation. On the other
hand, there will be special cases of higher symmetry (for
example a $W_3$ symmetry) for which one needs to include
a higher-spin gauging as well. 
If the higher symmetry is generated by 
%
holomorphic/antiholomorphic
polynomials $P_r(x^i,\diffp x^i)$, $\bar{P}_r(x^i,\diffm x^i)$,
the action has the form (\ref{yeq54}),
\bss
\bea
\label{eeq2}
S& =& \int d^2 \xi\, \cl
\\{}
\cl & = & \cl_G + \ppf [ \sum_r \alpha_r P_r(x^i,B^i) +
\sum_r \bar{\alpha}_r \bar{P}_r(x^i,\bar{B}^i)
\nonu \\{}
& & \qquad \qquad
 \quad -(B^i-\diffp x^i) G_{ij} (\bar{B}^j - \diffm x^j) ]
\label{eeq4}
\eea \label{eeeq}
\ess
where $\alpha \lt_{ij}
B^i B^j$ has been
replaced by $\sum_r \alpha_r P_r(x^i,B^i)$, and
similarly for the term involving $\bar{\alpha}$. 
These actions include the spin-two gauged sigma models
(\ref{e71}) when the generating polynomials $P,\bar{P}$ are
the $K$-conjugate stress tensors $8\pi \alpha'\tilde{T},
8 \pi \alpha' \tilde{\bar{T}}$. 
The action (\ref{eeeq}) is invariant under the 
higher-spin gauge
transformations
\bss\bea
& \delta \a_r  =  -\diffm \xi_r + \ldots & \\{}
& \delta x^i \equiv  \xi^i  =  \sum_r \xi_r G^{ij} 
\frac{\partial P_r(x^i,B^i)}{\partial B^j} &
\\{}
& \delta B^i  =  \diffp \xi^i - (B^j -\diffp x^j) \xi^k
 (\gh^+)_{jk}{}^i &
\\{}
& \delta \bb^i  =  -\xi^j \bb^k (\gh^-)_{jk}{}^i &
\eea\bea
& \bd \ba_r  =  -\diffp \bx_r + \ldots &
\\{}
& \bd x^i \equiv \bx^i  =
\sum_r \bx_r G^{ij} 
\frac{\partial \bar{P}_r(x^i,\bb^i)}{\partial \bb^j} &
\\{}
& \bd B^i  =  -\bx^j B^k (\gh^+)_{jk}{}^i &
\\{}
& \bd \bb^i  =  -\diffm \bx^i - (\bb^j - \diffm x^j)
\bx^k (\gh^-)_{jk}{}^i   &
\eea\ess
where the detailed form of $\delta \a_r$ and $\delta \ba_r$ depends
on the Poisson bracket algebra formed by $P_r$ and $\bar{P}_r$.
Adding to (\ref{eeeq}) an additional $B\bar{B}$-term,
one can also include a vector gauging
of non-abelian spin 1 symmetries.

It is quite
remarkable that by introducing auxiliary fields this large class of
actions can be brought to this simple polynomial form. Integrating
out the auxiliary fields yields the non-linear
form of these actions, which are also non-local in the general case.

\noindent
{\bf Acknowledgements}

We thank N. Obers, K. Sfetsos and A. Tseytlin for helpful discussions.
This research is supported in part by
NSF grant PHY-95-14797 and DOE grant DE-AC03-76SF00098.
JdB is a fellow of the Miller Institute for Basic Research in
Science.

\appendix{Sigma Model Form of the Spin-One Gauged
WZW Action}

The results of this appendix were worked out with K. Sfetsos.

The WZW action \cite{r20,r21} is 
\bea
 & S_g = -\frac{1}{8\pi y} 
\int d^2 \xi \tr( g^{-1} \diffp g g^{-1} \diffm g) -
\frac{1}{12\pi y} \int \tr(g^{-1} d g)^3  & \nonu \\{}
& \tr(T_a T_b) = y G_{ab}, \qquad
\diff_{\pm} = \diff_{\tau} \pm \diff_{\sigma}, \qquad
d^ 2 \xi = d\tau d\sigma
 \label{eq145} &
\eea
and the gauged WZW action \cite{r22,r23,kahs,kas}, which describes the
$g/h$ coset constructions \cite{r6,r10,r11}, is
\be
\label{eq146}
S_{g/h}=S_g + \frac{1}{4\pi y} \int d^2 \xi \tr [
ig^{-1} \diffp g A_- -iA_+ \diffm g g^{-1}  - g^{-1} A_+ g A_- + A_+ A_-] .
\ee
The equations of motion of the gauge fields $A_{\pm}$ are
\bea
\label{eq147}
&& g^{-1} D_+ g |_h = D_- g g^{-1}|_h =0 \nonu \\{}
&& D_{\pm} g = \diff_{\pm} g + i[A_{\pm},g]
\eea
and the matter equations of motion can be broken apart to
read
\bea
\label{eq148}
&& F_{-+} = \diffm A_+ - \diffp A_- + i[A_-,A_+] =0
\nonu \\{}
&& D_-(g^{-1} D_+ g) =0 .
\eea
To go to the sigma model form of the coset actions, we gauge fix the $h$
invariance of (\ref{eq146}), integrate out $A_{\pm}$ and compare
to the sigma model (\ref{u51}) and its equations of motion (\ref{u53}). This 
gives the coset metric, vielbein and spin connections in
(\ref{u11}).

Moreover, we find for the gauge fields that
\bea
\label{eq149}
(A_+)_a{}^b & = & A_+^A (-if_{Aa}{}^b ) = i\diffp x^i 
(\hat{\omega}^-_i(\Lambda))_a{}^b \nonu \\{}
(A_-)_a{}^b & = & A_-^A(-i f_{Aa}{}^b) = i\diffm x^i (\omhp_i)_a{}^b
\eea
where $\hat{\omega}^+$ and
$\hat{\omega}^-(\Lambda)$ are given in (\ref{eq111}) and 
(\ref{eq113}). Using these
relations, the flatness condition in (\ref{eq148}) can be
expressed as 
\bea
\label{eq150}
\hat{R}^+_{cda}{}^b + \delhp_d \Phi_{ca}{}^b & = & 0 \nonu \\{}
(\Phi_i)_a{}^b & \equiv & (\omhp_i)_a{}^b - \omhm_i(\L)_a{}^b \nonu \\{}
& = & (1+M+M^T)^A{}_{\mu} L_i{}^{\mu} f_{Aa}{}^b
\eea
where the matrix $M$ is given in
(\ref{eq109}) and 
$\hat{R}^+_{cda}{}^b$ is the generalized Riemann tensor with torsion
(see \cite{r14} and Section~2). The sigma model Einstein equations are 
\be
\hat{R}^+_{ij} = 2\delhp_i \delhp_j \Phi
\label{eq151}
\ee
where $\Phi$ is the dilaton. Comparing (\ref{eq150}) and
(\ref{eq151}) gives the known \cite{r24} form of the coset dilaton 
\bss\bea
\Phi& = & \frac{1}{2} \log\det (P_h N P_h) =
-\frac{1}{2} \log\det (P_h(1-P_h\Omega P_h) P_h)
\label{eq152}
\\{}
2\diff_i \Phi &  = &  (\hat{\omega}^+_a - \omhm_a(\L))_i{}^a =
e_i{}^a M_b{}^A f_{Aa}{}^b
\label{eq153}
\eea\ess
in a new way. We found that the identities
\bea
(1+M^T) P_{g/h} (1+M) & = & 1+M+M^T
\label{n80} \\{}
M^a{}_A M_b{}^B f_{Ba}{}^b & = & f_{BA}{}^D M_D{}^B
\label{n81}
\eea
were helpful in the algebra above.

\appendix{Semiclassical Limit of the Lie $h$-Invariant CFTs}

Here, we review and extend some facts \cite{r16} about the local
Lie $h$-invariant CFTs.

The general affine-Virasoro construction on simple $g$ is \cite{r1,r2}
\bss\bea
\label{eq154}
& T=L^{ab} :J_aJ_b: &
\\{}
& L^{ab} = 2 L^{ac} G_{cd} L^{db} - L^{cd} L^{ef} f_{ce}{}^a  f_{df}{}^b - 
L^{cd} f_{ce}{}^f f_{df}{}^{(a} L^{b)e} &
\label{eq155}
\\{}
\label{eq156}
& c=2G_{ab} L^{ab} &
\\{}
& G_{ab} = k\eta_{ab} &
\label{eq158}
\\{}
& a,b,c=1,\ldots, \dim\, g &
\label{eq157}
\eea\ess
where $a,b \ra \mu,\nu$ in the text, and (\ref{eq155}) is the
Virasoro master equation. The high-level smooth solutions
of the master equation have the semiclassical form
\be
\label{eq159}
L^{ab} = \frac{P^{ab}}{2k} + {\cal O}(k^{-2}), \qquad
P_a{}^c P_c{}^b = \d_a{}^b
\ee
where $P_a{}^b=G_{ac} P^{cb}$.

The Lie $h$-invariant CFTs on $g$, with $g/h$ a reductive
coset space, satisfy
\bss\bea
\label{eq160}
& a=(A,I),  \qquad A=1,\ldots,\dim\, h, \qquad I=1,\ldots,\dim\, g/h & 
\\{}
\label{eq161}
& G_{AI} = f_{AB}{}^I = f_{AI}{}^B =0 & 
\\{}
\label{eq162}
& L^{c(a} f_{cA}{}^{b)}=0 & 
\eea\ess
where $I,J \ra a,b$ in the text. The Lie $h$-invariant CFTs
fall into two classes, global and local,
according to the realization (\ref{eq163}), (\ref{eq164})
 of the $h$ symmetry.

It was observed \cite{r16} in the graph-theory ansatz on
$g=SO(n)$ that the local Lie $h$-invariant CFTs have
the semiclassical behavior
\be
L^{IJ} = {\cal O}(k^{-1}), \qquad
L^{AB}={\cal O}(k^{-2})
, \qquad L^{AI}= {\cal O}(k^{-2})
\label{eq167}
\ee
so that the coset-valued $L$ is semiclassically dominant.
It was argued in the text that this is true in general for
all local Lie $h$-invariant CFTs.

Then the master equation (\ref{eq155}) gives the leading
semiclassical form of the general local Lie
$h$-invariant CFT:
%
%
\bss\bea
\label{eq165}
& L^{IJ} = \frac{P^{IJ}}{2k} + {\cal O}(k^{-2}),
\qquad P^{IK} P_{K}{}^J = \d_I{}^J &
\\{}
\label{eq166}
& P^{K(I} f_{KA}{}^{J)} = 0 &
\\{}
%
\label{eq168}
& c={\rm rank}(P^{IJ}) + {\cal O}(k^{-1}) &
\\{}
\label{eq169}
& L^{AB} = -\frac{1}{4k^2} P^{IJ} P^{KL} 
f_{IK}{}^A f_{JL}{}^B +
{\cal O}(k^{-3}) &
\\{}
\label{eq170}
& L^{AJ}(\d_J{}^I - P_J{}^I) =
-\frac{1}{4k^2} P^{MN} (P^{JK} f_{JM}{}^A f_{KN}{}^I
 + f_{MJ}{}^L f_{NL}{}^A P^{IJ} ) 
+ {\cal O}(k^{-3}) .  & 
\eea\ess
It would be interesting to study the local Lie $h$-invariant CFTs 
in the one-loop unified Einstein-Virasoro master equation of
\cite{r14}, where the dilaton must simulate the $L^{AB}$ and
$L^{AI}$ contributions which are missing in the sigma model
description.

\end{document}